\documentclass[%
superscriptaddress,
 amsmath,amssymb,
 aps,
 prfluids,
 showkeys,
longbibliography 
]{revtex4-2}

\paperheight\pdfpageheight 
\paperwidth\pdfpagewidth   
\usepackage[bookmarksnumbered,raiselinks,breaklinks]{hyperref}
\hypersetup{colorlinks,%
    citecolor=blue,%
    filecolor=blue,%
    linkcolor=blue,%
    urlcolor=blue}
\usepackage{graphicx}
\usepackage[dvipsnames]{xcolor}
\usepackage{amsmath}
\usepackage[utf8]{inputenc}
\usepackage{natbib}

\def\clap#1{\hbox to 0pt{\hss#1\hss}}

\newcommand{\vect}[1]{\boldsymbol{#1}} 
\newcommand{\bnabla}{\vect{\nabla}}    
\newcommand{\nablaperp}{\nabla_{\!\!\perp}}    
\newcommand{\bnablaperp}{\vect{\nablaperp}}    
\newcommand{\intd}{\mathrm{d}}         
\newcommand{\I}{\mathrm{i}}            
\newcommand{\nondim}[1]{\tilde{#1}}    
\newcommand{\Real}{\mbox{\rm Re}} 
\newcommand{\Imag}{\mbox{\rm Im}} 
\newcommand{\hzero}{H_{0}}
\newcommand{\hinfty}{H_\infty}
\newcommand{\hzeroprime}{H_{0}'}
\newcommand{\vortzero}{\Xi_{0}}

\begin{document}

\title{A deep-water closure model for surface waves on axisymmetric swirling flows}

\author{Emanuele Zuccoli}
\affiliation{Mathematics Institute, University of Warwick, Coventry, CV4 7AL, United Kingdom}
\affiliation{School of Mathematics, Cardiff University, Cardiff, CF24 4AG, United Kingdom}

\author{Edward J.~Brambley}
\affiliation{Mathematics Institute, University of Warwick, Coventry, CV4 7AL, United Kingdom}
\affiliation{WMG, University of Warwick, Coventry, CV4 7AL, United Kingdom}

\author{Dwight Barkley}
\affiliation{Mathematics Institute, University of Warwick, Coventry, CV4 7AL, United Kingdom}

\date{\today}

\begin{abstract}
  {\centering%
    \href{https://doi.org/10.1103/PhysRevFluids.10.024801}{Published in Phys. Rev. Fluids 10, 024801, doi:10.1103/PhysRevFluids.10.024801}.%
    \\[1em]}

  We consider the propagation of linear gravity waves on the free surface of steady, axisymmetric flows with purely azimuthal velocity. We propose a two-dimensional set of governing equations for surface waves valid in the deep-water limit. 
These equations come from a closure condition at the free surface that reduces the three-dimensional Euler equations in the bulk of the fluid to a set of two-dimensional equations applied only at the free surface.
Since the closure condition is not obtained rigorously, it is validated numerically through comparisons with full three-dimensional calculations for vortex flows, including for a Lamb--Oseen vortex.
The model presented here overcomes three limitations of existing models, namely: it is not restricted to potential base flows; it does not assume the base flow to have a flat free surface; and it does not require the use of infinite-order differential operators (such as $\tanh(\nabla)$) in the governing equations. 
The model can be applied in the case of rapid swirl (large Froude number) where the base free surface is substantially deformed.  
Since the model contains only derivatives of finite order, it is readily amenable to standard numerical study. 

\end{abstract}


\maketitle


\section{Introduction}

Interaction between waves and vortices has long been a source of study in different fields of fluid dynamics, not only due to the beauty of these phenomena, but also owing to the rich physics and variety of applications they have. Examples can be found in  
acoustics~\citep{fetter,nazareko+zabusky+scheidegger,kopiev+belyaev}, geophysical fluid dynamics
\citep{buhler+mcintyre} and wave generation by turbulence \citep{lund,cerda+lund}. 
In this paper, we are interested in the interaction between surface waves and free-surface swirling flows, i.e. vortical flows whose upper boundary is not fixed, but is delimited by the free surface of the fluid. The corresponding linear perturbation problems are in principle three dimensional (3D) in space, giving rise to computationally costly problems to be solved. 
The authors previously studied~\citep{zuccoli+brambley+barkley-2022} the 3D linear perturbation problem for a Lamb--Oseen base flow in a horizontally unbounded region with a finite, non-small depth, using a modal analysis. These computations show the difficulty in resolving just normal mode eigenfunctions for extreme parameter values, let alone the difficulties of performing simulations of general time-dependent linear waves over strong vortices in deep water.
In addition to dealing with the deformed free surface, a substantial difficulty is the need to deal with effectively unbounded horizontal directions, which necessitates implementing some method to eliminate wave reflections at computational boundaries. This can be both difficult and computationally costly.
In order to partially overcome such difficulties, 
simple swirling flows, such as solid-body rotation and a potential vortex in a finite cylindrical container, have been considered in \citep{acheson,tophoj+others,jansson+others,bach+others,mougel-2014,mougel-2015,mougel-2017} where the full perturbation equations, including axial dependence, could be addressed using modest computational resources.
Otherwise, due to the numerical expense, waves on more complicated swirling flows have mostly been studied under the shallow-water approximation, resulting in a two-dimensional (2D) problem to solve.

The shallow-water equations were first derived by~\citet{saintvenant} in order to describe the time-dependent evolution of a fluid free surface without needing to solve the complete Navier-Stokes equations in the bulk of the fluid. 
The shallow-water approximation holds at wavelengths much longer than the fluid depth, but fails rapidly once this condition is not met, and results in non-dispersive waves where all frequencies travel at the same wave speed. 
A first attempt to include the effect of depth, and the associated dispersion in water waves, was carried out by~\citet{boussinesq-1871}, who derived a dispersive wave equation taking into account the effects of depth. 
The Boussinesq equation and subsequent models are all based on the potential flow assumption, as well as the approximation that the free surface is flat and is located at a level, say $z = H$, above a rigid base at $z=0$. 
One strategy for deriving  Boussinesq-type equations is to depth average the governing Laplace equation for the velocity potential and then to use integration by parts to link the average potential and the potential evaluated on the free surface. By doing so, one can obtain a hierarchy of wave equations that approximate with increasing accuracy the dispersive character of the surface waves propagating on a finite-depth fluid. 
The accuracy of the model 
is dictated by the truncation in the series of integrals computed. In particular, Boussinesq kept only the first three terms in the series coming from integration by parts, obtaining the wave equation
\begin{equation}
\begin{aligned} &
\left(\!1 - \frac{H^2}{2}\nablaperp^2\!\right)\!\frac{\partial^2 \phi}{\partial t^2} - gH\!\left(\!1 - \frac{H^2}{6}\nablaperp^2\!\right)\!\nablaperp^2 \phi = 0,
\end{aligned}
\label{equ:bousinnesq_eq_final_in_2D}
\end{equation} 
where $\phi$ is the velocity potential on the free surface $z=H$ and $\bnablaperp = (\partial_{x}, \partial_{y})$ is the horizontal gradient operator. 
Recent, more accurate models have been derived by, in effect, keeping higher-order terms in the series.
The pseudo-differential equation obtained by 
\citet{milewski+keller},
\begin{equation}
\begin{aligned} &
\frac{\partial^2 \phi}{\partial t^2} - \I g\nablaperp\tanh(-\I H\nablaperp)\phi = 0,
\end{aligned}
\label{equ:infinite_bousinnesq_eq_final_in_2D_short_form_final}
\end{equation} 
may be viewed as resulting from retaining all terms in such a series expansion. Equivalently, this equation is obtained by the Dirichlet-Neumann operator for the velocity potential at the free surface \cite{craig1993numerical,craig1994hamiltonian}.
The operator $-\I\nablaperp\tanh(-\I H\nablaperp)$ can easily be defined in Fourier space, leading to the familiar dispersion relation $\omega^2=g|k|\tanh(|k|H)$, but contains infinitely many spatial derivatives which complicates its numerical solution in the spatial domain.

To study surface waves on a vortex, \citet{torres+coutant+dolan+weinfurtner} generalized Eq.~\eqref{equ:infinite_bousinnesq_eq_final_in_2D_short_form_final} to linear waves on a free surface over a background potential flow denoted $\bf U_0$.
By assuming that the background free surface remains flat, they obtained the equation
\begin{equation}
\begin{aligned} &
D^{2}_{t} \phi - \I g\nablaperp\tanh(-\I H\nablaperp)\phi = 0,
\end{aligned}
\label{equ:infinite_bousinnesq_eq_final_in_2D_short_form_final_conv}
\end{equation}
where $\phi$ is again the velocity potential for linear perturbations, now with respect to a non-zero background flow, and $D_t = \partial_t + \vect{U_0\! \cdot\! \nabla}$ is the convective derivative with respect to the background flow.
By means of ray-tracing \citep{buhler-2014}, \citet{torres+coutant+dolan+weinfurtner} 
studied equation (\ref{equ:infinite_bousinnesq_eq_final_in_2D_short_form_final_conv}) analytically giving predictions for both growth rates and oscillation frequencies of  
surface waves propagating over a potential vortex. 

In the present paper, in section~\ref{sec:maths_model} we propose a set of reduced model equations for linear surface waves in deep water, accounting for both a deformed mean free-surface and vorticity of the underlying background flow, without resorting to the infinite-order derivatives of \eqref{equ:infinite_bousinnesq_eq_final_in_2D_short_form_final}--
\eqref{equ:infinite_bousinnesq_eq_final_in_2D_short_form_final_conv}, thus allowing for numerical solution by straightforward methods.
The key point in our derivation is the introduction of a closure condition imposed along the background free surface in section~\ref{sec:closure}. 
We justify the validity of this closure condition in section~\ref{sec:validation} by means of numerical results and comparisons with full three-dimensional calculations from \citep[][]{zuccoli+brambley+barkley-2022} for three vortex profiles, including a Lamb--Oseen vortex.
Finally, conclusions and opportunities for future research are described in section~\ref{sec:conclusions}.

\section{Mathematical Model}\label{sec:maths_model}

\subsection{Governing equations and base flow}

We assume that viscosity is negligible and hence that the fluid flow is governed by the incompressible Euler equations,
\begin{align}
&\frac{\partial\vect{U}}{\partial t} + \vect{U\cdot \nabla}\vect{U} = - \frac{1}{\rho}\bnabla P  - g\vect{\hat{z}}, &&
\vect{\nabla \cdot U} = 0,
\label{equ:euler}
\end{align}
where $\vect{U}$ is the velocity, $P$ is the pressure, $\vect{\hat{z}}$ is a unit vector in the vertical direction, and the constants $\rho$ and $g$ are the fluid density and the acceleration due to gravity respectively.  
The fluid is contained between a bottom boundary at $z=0$ and an upper free surface at $z = H$, where in general $H$ varies in space. The fluid must satisfy no penetration through the bottom boundary, so that $\vect{U\cdot\hat{z}} = 0$ at $z=0$.
The flow must satisfy kinematic and dynamic boundary conditions at the free surface. We neglect surface tension and assume the fluid above the free surface to be dynamically passive, and in particular, to have a constant pressure $\bar{P}$.  Together, these give the boundary conditions
\begin{subequations}\label{equ:full_bcs}\begin{align}
\frac{\partial H}{\partial t} + \vect{U\cdot \nabla} H &= \vect{U\cdot\hat{z}}, &
&\text{and}&
P &= \bar{P}, &
\text{at} &\quad
z = H.
\tag{\theequation a,b}
\end{align}\end{subequations}

We consider any swirling base flow of the form $\vect{U}_{0}(r, \theta, z) = U_{0}(r)\vect{\hat{\theta}}$, where $(r,\theta,z)$ are the usual cylindrical coordinates, and such that $U_{0}(\infty) = 0$ so as the vortex decays at infinity. 
The governing equations and boundary conditions are satisfied by such a flow provided the base pressure and surface height are, respectively,
\begin{align}
P_0(r, z) &= \bar{P} + \rho g \left( \hzero(r) - z\right), &
\hzero(r) &= \hinfty - \frac{1}{g}\int_r^\infty \frac{U^{2}_{0}(\hat{r})}{\hat{r}}\,\intd\hat{r},
\end{align}%
\label{eq:P0_h0}%
where $\hinfty$ is the depth of the fluid at $r=\infty$.  The base flow angular velocity $\Omega_0(r) = U_0(r)/r$ and axial vorticity $\vortzero(r) = (rU_0(r))'\!/r$ will also be useful in what follows, where $'$ denotes differentiation with respect to the single argument $r$.

\subsection{Linearized equations}

We are interested in the behavior of infinitesimal perturbations to the base flow. 
Let $\vect{u}(r, \theta, z, t) = u_{r}(r, \theta, z, t)\vect{\hat{r}} + u_{\theta}(r, \theta, z, t)\vect{\hat{\theta}} + u_{z}(x, y, z, t)\vect{\hat{z}}$ be the perturbation velocity, $p(r, \theta, z, t)$ the perturbation pressure, and $h(r, \theta, t)$ the perturbation height, so that the total velocity is $\vect{U} = U_0\vect{\hat{\theta}} + \vect{u}$, the total pressure is $P = P_0 + p$, and the total fluid height is $H = H_0 + h$. Assuming the perturbations to be small, the perturbation dynamics are given by linearizing the incompressible Euler equations~\eqref{equ:euler} and boundary conditions~\eqref{equ:full_bcs} about the base flow:
\begin{subequations}\label{linear_euler_equations_cartesian}
\begin{align} &
D_{t}u_{r} -2\Omega_0 u_{\theta} + 
\frac{1}{\rho}\frac{\partial p}{\partial r} = 0, \label{eq:linur} \\ &
D_{t}u_{\theta} + \vortzero u_r + 
\frac{1}{\rho r}\frac{\partial p}{\partial \theta} = 0, \label{eq:linut} \\ &
D_{t}u_{z} + \frac{1}{\rho}\frac{\partial p}{\partial z} = 0, \label{eq:linuz} \displaybreak[0]\\ &
\frac{1}{r}\frac{\partial \big( r u_r \big)}{\partial r} + \frac{1}{r}\frac{\partial u_\theta}{\partial\theta} + \frac{\partial u_z}{\partial z} = 0, \label{eq:div} \displaybreak[0]\\ &
u_{z} = 0, \quad {\rm on} \quad z = 0,  \label{eq:bottomBC}\\ &
D_{t}h -u_{z} + \hzeroprime u_{r} = 0, \quad {\rm on} \quad z = \hzero(r),
\label{eq:kinBC}\\ &
p - \rho gh = 0, \quad {\rm on} \quad z = \hzero(r),\label{eq:pBC}
\end{align}%
\label{eq:lin_full}%
\end{subequations}%
where $D_t = \partial/\partial t + \vect{U_0\cdot\nabla} \equiv \partial/\partial t + \Omega_0(r) \partial/\partial \theta $ is the convective derivative with respect to the base flow, as before, and 
primes denote differentiation with respect to the single coordinate $r$, as before. 
Note that equation~\eqref{eq:pBC} above is the linearization of the dynamic boundary condition (\ref{equ:full_bcs}b) and enforces that the total pressure $P_0+p$ on the free surface is atmospheric pressure.
While we cannot introduce a velocity potential here as our flow is not in general irrotational, for irrotational flows Bernoilli's equation would give $p = -\rho D_t\phi$ where $\phi$ is the perturbation velocity potential, and so equation~\eqref{eq:pBC} could then be written involving a temporal derivative of a velocity potential, which may be more familiar to some readers.

Equations~\eqref{eq:lin_full} apply for any type of linear perturbation to the base flow. Our interest here is in surface waves in deep water, for which we expect exponential decay of the perturbations away from the free surface and hence that the bottom boundary condition, Eq.~\eqref{eq:bottomBC}, is unimportant~\citep[see e.g.][pp. 63--65]{johnson-1997}. We define the unknowns on the base free surface as 
\begin{align}
u &= u_{r}|_{\hzero}, & v &= u_{\theta}|_{\hzero}, & w &= u_{z}|_{\hzero}, & h &= \frac{1}{\rho g}p|_{\hzero},
\end{align}
where here and throughout $|_{\hzero}$ means evaluated at $z = \hzero(r)$. The expression for $h$ is not a new definition, but is rather a consequence of the dynamic boundary condition \eqref{eq:pBC}. The model will be expressed in these free-surface unknowns.

We will use some identities for derivatives of quantities evaluated on the surface.  Let $f(r, \theta, z, t)$ represent any of the four unknowns $u_r$, $u_\theta$, $u_z$ or $p$. Then, 
\begin{subequations}\begin{gather}
\frac{\partial (f|_{\hzero})}{\partial r} =
\left.\frac{\partial f}{\partial r}\right|_{\hzero}
+ \hzeroprime\left.{\frac{\partial f}{\partial z}}\right|_{\hzero}, \\
D_{t}\left(f|_{\hzero}\right) = \left.{(D_{t}f)}\right|_{\hzero} + \left(\boldsymbol{U}_{0}\cdot\vect\nabla \hzero\right)\left.{\frac{\partial f}{\partial z}}\right|_{\hzero} = 
\left.{(D_{t}f)}\right|_{\hzero}\label{eq:Dtsurface}.
\end{gather}\label{equ:relations_unknwons_on_FS}\end{subequations}
The final equality holds because 
$\boldsymbol{U}_0$ is azimuthal while $\hzero$ varies only with $r$.
Using these identities~\eqref{equ:relations_unknwons_on_FS}, the linearized momentum equations
\eqref{eq:linur}-\eqref{eq:linuz}, the incompressibility constraint \eqref{eq:div}, and the kinematic boundary condition \eqref{eq:kinBC}
evaluated on the free surface $z=\hzero(r)$ become
\begin{subequations}
\label{linearized_Euler_2D_base_flow_on_FS}
\begin{align} &
D_{t}u -2\Omega_0 v
+ g\frac{\partial h}{\partial r} - \frac{\hzeroprime}{\rho}\!\left.{\frac{\partial p}{\partial z}}\right|_{\hzero} \!\!\!= 0, \label{equ:x_mom_eq_on_FS1}\\ &
D_{t}v + \vortzero u
+ \frac{g}{r}\frac{\partial h}{\partial \theta} = 0, \label{equ:y_mom_eq_on_FS1}\\ &
D_{t}w + 
\frac{1}{\rho}\!\left.{\frac{\partial p}{\partial z}}\right|_{\hzero} \!\!\!= 0, \label{equ:z_mom_eq_on_FS1}\displaybreak[0]\\ &
\frac{1}{r}\frac{\partial (ru)}{\partial r} + \frac{1}{r}\frac{\partial v}{\partial \theta} + 
\!\left.{\frac{\partial u_{z}}{\partial z}}\right|_{\hzero}\!\!\!
- \hzeroprime\!\left.{\frac{\partial u_{r}}{\partial z}}\right|_{\hzero}\!\!\! = 0, \label{equ:cont_eq_on_FS1}\\ &
D_{t}h - w + \hzeroprime u = 0.
\label{equ:FS_BC}
\end{align}
\end{subequations} 
Exploiting equations \eqref{equ:z_mom_eq_on_FS1} and \eqref{equ:FS_BC}, $w$ and $\partial_z p$ can be eliminated from the system and the remaining equations become
\begin{subequations}
\begin{align} &
D_{t}u - 2\Omega_0 v
 + g\frac{\partial h}{\partial r} + \hzeroprime D_{t}\!\left( 
D_{t}h + \hzeroprime u \right) = 0,  \\ & 
D_{t}v + \vortzero u
 + \frac{g}{r}\frac{\partial h}{\partial \theta} = 0, \\ &
\frac{1}{r}\frac{\partial (ru)}{\partial r} + \frac{1}{r}\frac{\partial v}{\partial \theta} + 
\!\left.{\frac{\partial u_{z}}{\partial z}}\right|_{\hzero}
- \hzeroprime\!\left.{\frac{\partial u_{r}}{\partial z}}\right|_{\hzero}\!\!\! = 0. \label{equ:cont_eq_on_FS1_b}
\end{align}%
\label{x_and_y_momentum_and_cont_eqs_on_FS}%
\end{subequations}
Equations~\eqref{x_and_y_momentum_and_cont_eqs_on_FS} are expressed entirely in terms of the free-surface unknowns $u, v, h$, with the exception of the last two terms in the continuity equation~\eqref{equ:cont_eq_on_FS1_b}. 
To express these terms as function of $u, v$ and $h$ we next consider a closure model.

\subsection{Closure condition}
\label{sec:closure}

The closure for the continuity equation~\eqref{equ:cont_eq_on_FS1_b} will take the form of a boundary condition imposed along the base free surface. 
We begin by defining the functional
\begin{align} &
    \Psi[u_{r}, u_{z}, p](r, \theta, z, t) = u_{z} - \hzeroprime u_{r} - \frac{1}{\rho g}D_{t}p,
\label{functional_Psi}
\end{align}
such that kinematic boundary condition \eqref{eq:kinBC} is obtained by setting the functional to zero
along the free surface: $\Psi(r, \theta, \hzero(r), t) = 0$.
The ansatz we make to close the system is that the axial gradient of $\Psi$ evaluated on the free surface is also zero:
\begin{align}
\left.{\frac{\partial \Psi}{\partial z}}\right|_{\hzero}\!\!\! = \left.{\frac{\partial u_{z}}{\partial z}}\right|_{\hzero}\!\!\! - \hzeroprime\!\left.{\frac{\partial u_{r}}{\partial z}}\right|_{\hzero}\!\!\! 
- \frac{1}{\rho g}\!\left.{\frac{\partial (D_{t} p)}{\partial z}}\right|_{\hzero}\!\!\!= 0.
\label{differentiation_FSBC_2}
\end{align}
From this ansatz, we obtain 
\begin{subequations}
\begin{eqnarray}
  \left.{\frac{\partial u_{z}}{\partial z}}\right|_{\hzero}\!\!\! - \hzeroprime\!\left.{\frac{\partial u_{r}}{\partial z}}\right|_{\hzero}\!\!\! 
  & = &
    \frac{1}{\rho g}D_{t}\!\left.{\frac{\partial p}{\partial z}}\right|_{\hzero} \label{eq:closure_BC_p} \\
  & = & -\frac{1}{g} D^{2}_{t}w \label{eq:closure_BC_w} \\
  & = & -\frac{1}{g}D^{2}_{t}\!\left(D_{t}h + \hzeroprime u \right), \label{eq:closure_BC_final}
\end{eqnarray}%
\label{eq:closure_BC_all}%
\end{subequations}%
where the first line uses~\eqref{eq:Dtsurface}, the second line uses~\eqref{equ:z_mom_eq_on_FS1}, and the third line uses~\eqref{equ:FS_BC}.
While all three forms are equally valid, we will use the first for numerical verification (Sec.~\ref{sec:validate_closure}); the second form is most readily relatable to the simpler case of waves on a fluid at rest (Appendix~\ref{appendix:2D_closure}); the final form is used to close the system in terms of variables evaluated only on the free surface, as we now discuss.

Substituting~\eqref{eq:closure_BC_final} into the continuity equation \eqref{equ:cont_eq_on_FS1_b}, we arrive at our final set of governing equations, which are a 2D system of equations expressed in terms of the variables evaluated at the free surface, $u(r,\theta,t)$, $v(r,\theta,t)$, and $h(r,\theta,t)$,
\begin{subequations}
\begin{align} &
D_{t}u - 2\Omega_0 v
+ g\frac{\partial h}{\partial r} + \hzeroprime D_{t}\!\left( 
D_{t}h + \hzeroprime u \right) = 0, \\ &
D_{t}v + \vortzero u
+ \frac{g}{r}\frac{\partial h}{\partial \theta} = 0, \\ &
\frac{1}{r}\frac{\partial (ru)}{\partial r} + \frac{1}{r}\frac{\partial v}{\partial \theta} 
-\frac{1}{g}D^{2}_{t}\!\left(D_{t}h + \hzeroprime u \right) = 0.
\end{align}
\label{final_reduced_system}
\end{subequations} 

Ansatz~\eqref{differentiation_FSBC_2} has been assumed above without justification; while we have been unable to derive it from the full governing equations under suitable assumptions, and so we do not claim it to be exactly true (even in an asymptotic sense), we show below in section~\ref{sec:validation} that it is nonetheless satisfied to a high degree of accuracy for surface waves calculated numerically using the full three-dimensional governing equations. Some insight into the ansatz and closure can be found in Appendix~\ref{appendix:2D_closure}. There we elaborate on the parallel between the model equations for the swirling 3D base flow and those that are obtained in the same way for a simple 2D fluid. This suggests an alternative view towards deriving the model. Rather than invoking ansatz~\eqref{differentiation_FSBC_2}, we may assume that \eqref{eq:closure_BC_w} holds. This equation is a natural generalization of the case for a 2D fluid (see Appendix~\ref{appendix:2D_closure}). 
If Eq.~\eqref{eq:closure_BC_w} holds then all equalities in Eqs.~\eqref{eq:closure_BC_all} hold.
Beyond this analogy, we have been unsuccessful in our attempts to rigorously or formally justify \eqref{differentiation_FSBC_2} 
~\citep{zuccoli-2023}.
We therefore resort to numerical evidence for the validity of the closure.
In Sec.~\ref{sec:validate_closure}, we validate closure condition \eqref{eq:closure_BC_final} for linear modes on the free surface of vortices in deep water. Then in Sec.~\ref{sec:eval_validation} we show that modal solutions of the reduced 2D equations \eqref{final_reduced_system} match very accurately modal solutions of the full 3D Euler equations \eqref{linear_euler_equations_cartesian}.

\subsection{Non-dimensionalization}

We express the equations in non-dimensional form. We assume the base flow is characterised by a reference velocity $\bar{U}$ and by a reference length $a$ and use the following non-dimensionalization
\begin{equation}
    \begin{aligned} &
        r=a\nondim{r}, \qquad \theta=\nondim{\theta}, \qquad t= \sqrt{\frac{a}{g}}\nondim{t}, \\ &
        (u, v)= \sqrt{ag}(\nondim{u}, \nondim{v}), \qquad h= a\nondim{h}, \\ &
        U_{0} = \bar{U}\nondim{U}_{0}, \qquad \hzero= a\nondim{\hzero}.
    \end{aligned}
\end{equation}
The resulting dimensionless equations read
\begin{subequations}
\label{final_system_single_vortex_dimensionless}
    \begin{align} &
    \left(1 + \nondim{{\hzeroprime}}^2\right)\nondim{D}_{t}\nondim{u} - 2F\nondim{\Omega}_{0}\nondim{v} + \frac{\partial \nondim{h}}{\partial \nondim{r}} + \nondim{\hzeroprime}\nondim{D}^{2}_{t} \nondim{h} = 0, \\ &
    \nondim{D}_{t}\nondim{v} + F\nondim{\vortzero}\nondim{u} + \frac{1}{\nondim{r}}\frac{\partial \nondim{h}}{\partial \nondim{\theta}} = 0, \\ &
    \frac{1}{\nondim{r}}\frac{\partial }{\partial \nondim{r}}\left( \nondim{r}\nondim{u} \right) + \frac{1}{\nondim{r}}\frac{\partial \nondim{v}}{\partial \nondim{\theta}} - \nondim{D}^{2}_{t}\left(\nondim{D}_{t}\nondim{h} + \nondim{\hzeroprime}\nondim{u}\right) = 0,
    \end{align}%
\label{eq:closure_cyl}%
\end{subequations}%
where $F = \bar{U}/\sqrt{ag}$ is the Froude number, $\nondim{\vortzero} = (\nondim{r}\nondim{U}_{0})'/\nondim{r}$ is the dimensionless base vorticity, $\nondim{\hzeroprime} = F^2 \nondim{r}\nondim{\Omega}_0$ is the dimensionless free surface deformation, and $\nondim{D}_{t} = \partial/\partial\nondim{t} + F\Omega_0\partial/\partial\nondim{\theta}$.
In what follows, all variables have been nondimensionalized, and we drop the tildes from nondimensional variables for readability.

For later validation of the closure condition, we will consider the non-dimensionalized version of \eqref{eq:closure_BC_p} 
given by
\begin{equation}\begin{aligned}
  \left.{\frac{\partial u_{z}}{\partial z}}\right|_{\hzero}\!\!\! - \hzeroprime\left.{\frac{\partial u_{r}}{\partial z}}\right|_{\hzero}\!\!\! &=  D_{t}\left.{\frac{\partial p}{\partial z}}\right|_{\hzero}.
\end{aligned}
\label{eq:closure_BC_final_monpolar_case}
\end{equation}

\section{Model validation}
\label{sec:validation}

We will numerically validate the model for surface gravity waves on axisymmetric vortices. While the model is a general time-dependent, 2D system of equations, we require solutions from the full 3D Euler equations with which to compare. For this we turn to modal solutions,
focusing primarily on the Lamb--Oseen vortex for which modal solutions of the Euler equations have been recently obtained~\cite{zuccoli+brambley+barkley-2022, zuccoli-2023}. 

We will first directly verify that, in the limit of deep water, eigenmodes for surface waves from the 3D linear Euler equations indeed satisfy closure condition
\eqref{eq:closure_BC_final_monpolar_case} to a high degree of accuracy. We will then show that solutions to the reduced 2D model \eqref{eq:closure_cyl} accurately reproduce surface gravity modes from the 3D Euler equations in deep water. 

\subsection{Numerical test case}

We will examine surface gravity waves on a base Lamb--Oseen vortex given by the azimuthal velocity and corresponding free-surface height
\begin{equation}
    U_0(r) = \frac{1 - \exp(-r^2)}{r}, \qquad \qquad
    \hzero(r) = \hinfty - F^2 \int_r^\infty \frac{U^{2}_{0}(r')}{r'}\,\intd r'.
\label{eq:Lamb_Oseen}
\end{equation}
This base flow has been non-dimensionalized by taking
size of the vortex core as the lengthscale $a$. 
The Froude number $F$ is a non-dimensional measure of the vortex circulation. The parameter $\hinfty$ is the non-dimensional fluid depth at infinity. Thus, the base flow is specified by two non-dimensional parameters: $F$ and $\hinfty$.

For full 3D Euler computations, we use the numerical method of \citet{zuccoli+brambley+barkley-2022}, concentrating on modal solutions.
Since the base flow is steady and axisymmetric, eigenmodes will have $(\theta,t)$ dependence of the form $\exp\{-\I\omega t + \I m\theta\}$, where $\omega$ is a (complex) temporal eigenvalue and $m$ is the (integer) azimuthal wavenumber.  
The linear Euler equations \eqref{linear_euler_equations_cartesian} can readily 
rendered into dimensionless form, and then transformed into an eigenvalue problem for $\omega$ and the corresponding eigenmodes. The eigenvalues and eigenmodes are in general complex, with $\Real(\omega)$ the frequency and $\Imag(\omega)$
the growth rate of the mode. The eigenvalue problem is discretized and solved numerically using a spectral method. 
To emulate an unbounded radial domain, an absorbing layer method is used to provide non-reflecting boundary condition at a finite spatial location $r = R \gg 1$. 
While we obtain all eigenvalues, we show below the dominant eigenvalues with the largest growth rate for specified values of $F, m$, and $\hinfty$. The corresponding eigenmodes are given by the fields $u_r(r,z)$, $u_\theta(r,z)$, $u_z(r,z)$, $p(r,z)$, and $h(r)$. See Zuccoli {\em et al.} \cite{zuccoli+brambley+barkley-2022, zuccoli-2023} for details. 

\subsection{Validating the closure condition}
\label{sec:validate_closure}

For the modal solutions, the convective derivative operator $D_{t} = \partial_{t} + F \Omega_{0}(r)\partial_{\theta}$ appearing in Eq.~\eqref{eq:closure_BC_final_monpolar_case} is given by $D_{t} = -\I\omega + \I m F \Omega_{0}(r)$. Hence, the closure condition that we will validate takes the form
\begin{equation}
  \left.{\frac{\partial u_{z}}{\partial z}}\right|_{\hzero}\!\!\! - \hzeroprime\left.{\frac{\partial u_{r}}{\partial z}}\right|_{\hzero}\!\!\! =  
  \big({-\I\omega} + \I mF\Omega_{0}\big)\!\left.{\frac{\partial p}{\partial z}}\right|_{\hzero}.
\label{eq:closure_BC_mode}
\end{equation}

We begin by examining one typical surface eigenmode. We set $F=0.5$ and $m=7$, and consider the least damped mode as we vary the fluid depth $\hinfty$.
\begin{figure}
\centering%
    \includegraphics[]{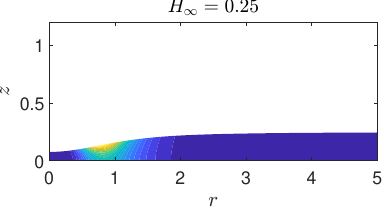}%
    \includegraphics[]{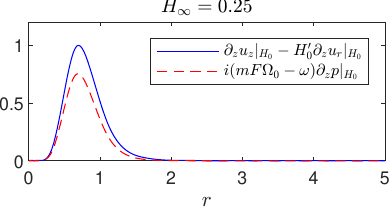}%
    \par\vspace{.4cm}
    \includegraphics[]{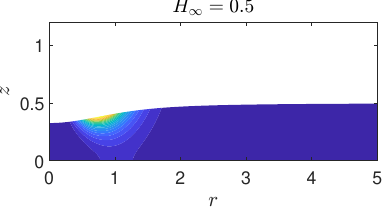}%
    \includegraphics[]{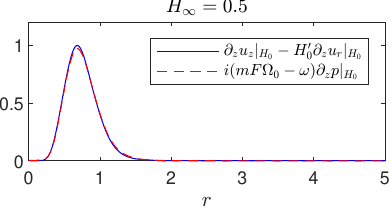}%
    \par\vspace{.4cm}
    \includegraphics[]{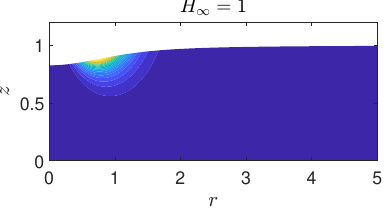}%
    \includegraphics[]{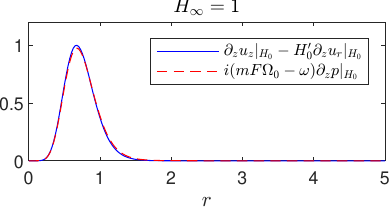}%
\caption{Accuracy of the closure condition for the least-damped surface-wave modes with Froude number $F = 0.5$ and azimuthal wavenumber $m = 7$ at three values of fluid depth. The three modes are all marginally stable with growth rate $\Imag(\omega) \sim -10^{-9}$. Left panels illustrate the modes with contours of the modulus of the pressure field $|p(r,z)|$. Contours are equally spaced with blue representing zero pressure magnitude and yellow representing the maximum pressure magnitude. The modulus of both left-hand and right-hand sides of the closure condition are evaluated and plotted in the right panels, with the eigenfunction normalized so that the maximum value of the left-hand side is $1$.}
\label{fig:closure_trapped}%
\end{figure}%
Figure~\ref{fig:closure_trapped} (left panels) illustrate the form of the eigenmodes, visualized using the modulus of the pressure field $|p(r,z)|$, at three values of $\hinfty$. 
The shape of the base free surface due to the base vortex is evident. The eigenmode is localized to the core region of the vortex and is referred to as a trapped mode. (The eigenmode ``seen from above'' is shown in figure~\ref{fig:commparison_3D_model_from_above} and is discussed below.) 
For the top results, at a depth of $\hinfty=0.25$, it is visually evident that the vertical structure of the mode is affected by the bottom boundary, and this is therefore an intermediate-depth case between deep water, where the mode is unaffected by the bottom boundary, and shallow water, where the mode is entirely independent of $z$. The bottom results at a depth of $\hinfty=1$ is comfortably in the deep water regime.

\begin{figure}
    \centering
    \includegraphics[]{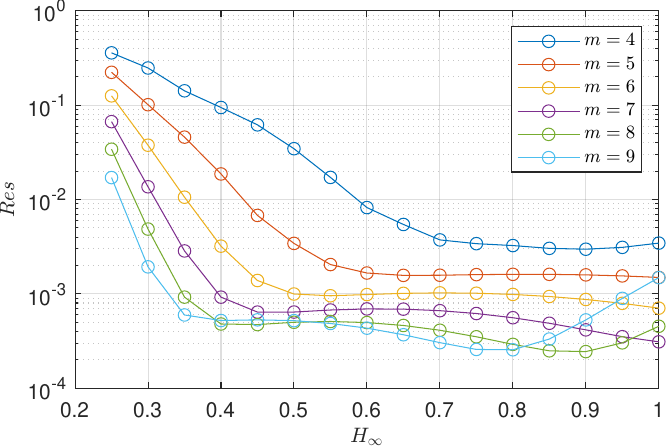}
    \caption{Trend of the residuals $Res = \|\mathrm{LHS}-\mathrm{RHS}\|_2/\|\mathrm{LHS}\|_2$ on a $\mathrm{log}$-scale as function of the fluid depth $H_\infty$ for the least damped modes computed at $F = 0.5$ and several azimuthal wavenumbers $m$.}
    \label{fig:residuals_various_m}
\end{figure}

To validate the closure condition, we compute the left-hand side and right-hand side of equation \eqref{eq:closure_BC_mode} from the eigenmode fields and plot these as a function of $r$ in the right panels of figure \ref{fig:closure_trapped}. Eigenmodes have been normalized so that the maximum value of the left-hand side is 1 in each case. 
While at $\hinfty = 0.25$ the closure condition is not satisfied, for depth $\hinfty = 0.5$ and larger the closure condition is well satisfied.  To quantify this, we compute the residual $Res = \|\mathrm{LHS}-\mathrm{RHS}\|_2/\|\mathrm{LHS}\|_2$, the $L_2$ norm of the difference between the left-hand-side ($\mathrm{LHS}$) and right-hand-side ($\mathrm{RHS}$) of equation~\eqref{eq:closure_BC_mode}, normalized by the $L_2$ norm of the left-hand-side, for each of the cases shown. The residual gives $Res = [6.7 \times 10^{-2},\, 1.2 \times 10^{-3},\, 8.7\times 10^{-4}]$ for $\hinfty = [0.25,\, 0.5,\, 1]$, respectively. 
More detailed results are shown in figure~\ref{fig:residuals_various_m}
where the residual for the least damped modes arising at $F = 0.5$ is plotted against the fluid depth $H_\infty$ for different azimuthal wavenumbers $m$. It can be noted that as the regime passes from shallow to deep water, the residuals decrease quickly (approximately exponentially) reaching a level of $10^{-3}$ which is close to the limits of what we can resolve with our numerical computations; the resolution used in the 3D computations has been chosen as a compromise between obtaining sufficiently well-resolved modes structure and computational time (see~\citep{zuccoli+brambley+barkley-2022} for details of the numerical method, including an extensive convergence study). The deeper the fluid and higher the azimuthal mode number, the thinner and more localized to the surface the structure of the modes becomes~\citep{zuccoli+brambley+barkley-2022}, thus the more difficult it becomes to properly compute and resolve those modes given a limited resolution. This explains the small increase in the residuals of figure \ref{fig:residuals_various_m} for $m = 8, 9$ close to $H_\infty = 1$.

We have conducted many similar tests of the closure condition for a variety of other surface waves and other base swirling flows, and in all cases we find that the closure condition is satisfied in deep water. Here, we present and discuss a few other illustrative cases.

\begin{figure}
\centering%
    \includegraphics[]{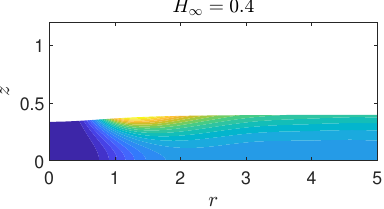}%
    \includegraphics[]{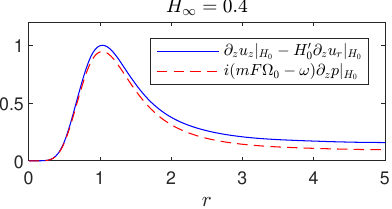}%
    \par\vspace{.4cm}
    \includegraphics[]{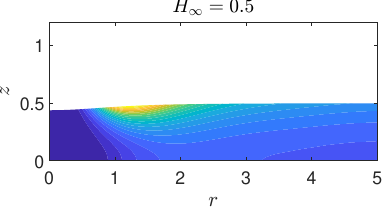}%
    \includegraphics[]{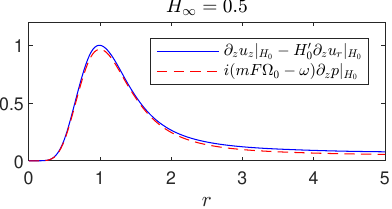}%
    \par\vspace{.4cm}
    \includegraphics[]{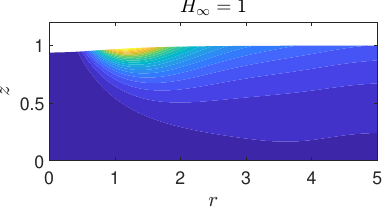}%
    \includegraphics[]{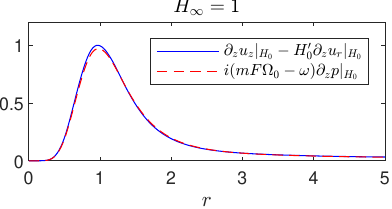}%
\caption{Accuracy of the closure condition for the least-damped surface-wave modes with Froude number $F = 0.3$ and azimuthal wavenumber $m = 7$ at three values of fluid depth. All three modes are nearly neutrally stable and are slowly decaying in time, with growth rates $\Im[\omega] = [-0.053, -0.031, -0.014]$, respectively. Left panels illustrate the modes with contours of the modulus of the pressure field $|p(r,z)|$. Contours are equally spaced with blue representing zero pressure magnitude and yellow representing the maximum pressure magnitude. The modulus of both left-hand and right-hand sides of the closure condition are evaluated and plotted in the right panels, with the eigenfunction normalized so that the maximum value of the left-hand side is $1$.}
\label{fig:closure_radiating}
\end{figure}%
Figure~\ref{fig:closure_radiating} shows a set of plots similar to those in figure \ref{fig:closure_trapped}, but at a smaller Froude number of $F=0.3$. In this case, the surface eigenmode is a radiating mode and extends radially outside the vortex core. (Figure
\ref{fig:commparison_3D_model_from_above} shows the eigenmode ``seen from above'' and is discussed below.) 
Compared with the previous case, here the base free surface is less deformed, the peak in the mode is shifted slightly to larger radius, and most significantly the horizontal lengthscale of the mode is larger. As a result, we observe that at $\hinfty = 0.4$ the mode is visibly distorted from the deep-water limit. Nevertheless, as the right-hand panels in figure \ref{fig:closure_radiating} show, the eigenmode fields satisfy the closure condition in deep water. Quantitatively, the residual norms in satisfying the closure condition for this case are $Res = [7.7\times 10^{-2},\, 4.3 \times 10^{-3},\, 8.2 \times 10^{-4}]$ for $\hinfty = [0.4,\, 0.5,\, 1]$, respectively.

\begin{figure}
\centering%
(a)%
    \raisebox{-0.9\height}{\includegraphics[]{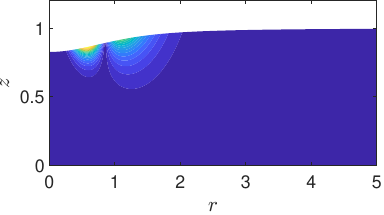}}%
    \raisebox{-0.9\height}{\includegraphics[]{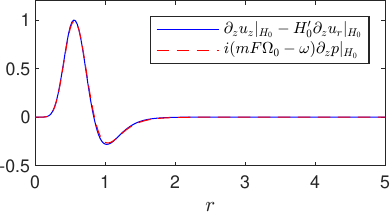}}%
    \par\vspace{.5cm}%
(b)%
    \raisebox{-0.9\height}{\includegraphics[]{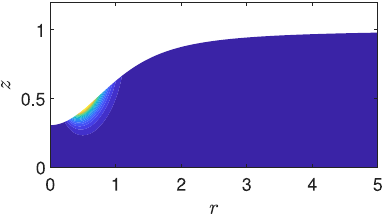}}%
    \raisebox{-0.9\height}{\includegraphics[]{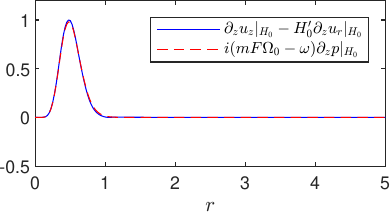}}%
    \par\vspace{.5cm}%
(c)%
    \raisebox{-0.9\height}{\includegraphics[]{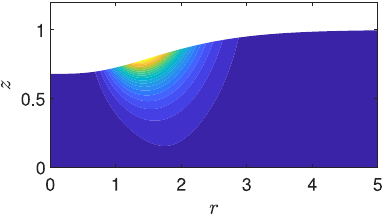}}%
    \raisebox{-0.9\height}{\includegraphics[]{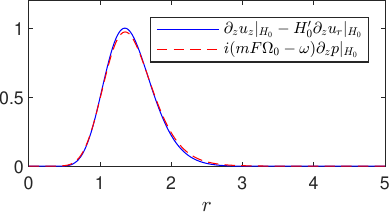}}%
    \par\vspace{.5cm}%
(d)%
    \raisebox{-0.9\height}{\includegraphics[]{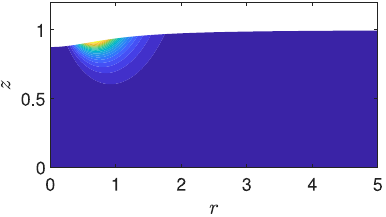}}%
    \raisebox{-0.9\height}{\includegraphics[]{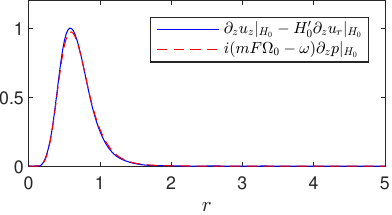}}%
    \par\vspace{.5cm}%
(e)%
    \raisebox{-0.9\height}{\includegraphics[]{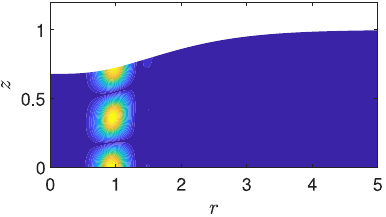}}%
    \raisebox{-0.9\height}{\includegraphics[]{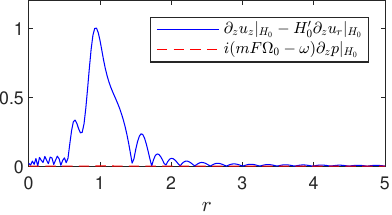}}%
\caption{(left) modulus of pressure eigenfunction in the $r$-$z$ plane.  (right) Modulus of the corresponding contributions to the closure condition.  Unless otherwise stated, $F=0.5$, $m=7$, $\hinfty=1$.
(a) A doubly-oscillatory surface wave mode, Lamb--Oseen profile, $\Imag(\omega) \sim -10^{-7}$.
(b) A surface wave mode, Lamb--Oseen profile, with $F = 1$, $\Imag(\omega) \sim -10^{-8}$.
(c) A surface wave mode, profile~(\ref{equ:Vortex2}), $\Imag(\omega) \sim -10^{-8}$.
(d) A surface wave mode, profile~(\ref{equ:BH_Vortex}), $\Imag(\omega) \sim -10^{-8}$. 
(e) An inertial wave mode, profile~(\ref{equ:Vortex2}), $\Imag(\omega) \sim -10^{-9}$.}%
\label{fig:other_vortices}%
\end{figure}%
Figure~\ref{fig:other_vortices} illustrates a variety of further cases with the same types of figures already shown. We forgo plotting different fluid depths and show only the deep-water cases. 
In figure~\ref{fig:other_vortices}(a), we show a higher-order surface gravity mode, comparable to that shown in the bottom plots of figure~\ref{fig:closure_trapped} but with two radial oscillations instead of one, and indeed the closure condition is satisfied in this case also.  This is typical of the agreement we have found for all surface gravity modes, irrespective of whether they are the least damped mode or not.
In figure \ref{fig:other_vortices}(b) we show a surface gravity mode at $F=1.0$ to illustrate a situation where the base free surface significantly deviates from flat. The closure condition is nevertheless satisfied by the surface gravity mode.

We have tested the accuracy of our closure for two other axisymmetric vortices given by the following angular-velocity profiles:
\begin{equation}
    \Omega_{0}(r) = \frac{r}{4}\exp\{ 2 - r \},
    \label{equ:Vortex2}
\end{equation}
and
\begin{equation}
    \Omega_{0}(r) = \frac{1}{r^2 + 1}.
    \label{equ:BH_Vortex}
\end{equation}
This second profile was considered by \citet{patrick-2018}. The results are shown in figures \ref{fig:other_vortices}(c) and \ref{fig:other_vortices}(d). The shape of the base free surface and the details of the surface gravity modes differ from those of the Lamb-Oseen vortex, but again the eigenmodes satisfy the closure condition. 

Finally, the model we derived applies to surface gravity modes and not the other type of linear waves that can arise in swirling flows, such as inertial or Rossby modes (see for example \citet{mougel-2015}, \citet{mougel-2017}). Thus, we conclude this section by showing explicitly that our closure condition fails for an inertial mode. Figure \ref{fig:other_vortices}(e) shows a representative inertial mode arising from vortex in equation \eqref{equ:Vortex2}, for $m = 7$ and $F = 0.5$. The closure condition fails to hold for such a mode. Indeed inertial modes arise when there is no momentum flux across the vertical boundaries \citep{greenspan-1969} and only the rotation of the base flow is responsible for these waves. This implies $\partial_z p|_{\hzero} \simeq 0$ on the free surface for the inertial mode. The right-hand side of the closure condition is zero while the left-hand side is not, giving a residual of $Res = 0.9975$.

\subsection{Comparison between full and model eigenmodes}
\label{sec:eval_validation}

This section is devoted to a side-by-side comparison of eigenmodes obtained from the linearized 3D Euler equations and those obtained from the reduced 2D system coming from the closure condition. 
The reduced 2D equations~\eqref{eq:closure_cyl}
can be re-expressed as an eigenvalue problem, and due to the modal dependence in the azimuthal direction $e^{im\theta}$, numerical computations are required to compute fields depending only on the radial coordinate $r$. 
Recall that for the Euler equations, the corresponding numerical computations must capture fields in a $(r,z)$-domain with a deformed free-surface, as in figures~\ref{fig:closure_trapped}-\ref{fig:other_vortices}.
Hence, the model both reduces the number of space dimensions and removes the need to compute fields in the complex regions bounded by the deformed base free surface. This is a significant saving of effort. On a more technical side, a further advantage of the model is that one can derive a non-reflecting boundary condition for the model as discussed in Appendix~\ref{appendix:model_NRBC}.

\begin{figure}
\centering%
    \par\vspace{.5cm}
    \includegraphics[width=5cm]{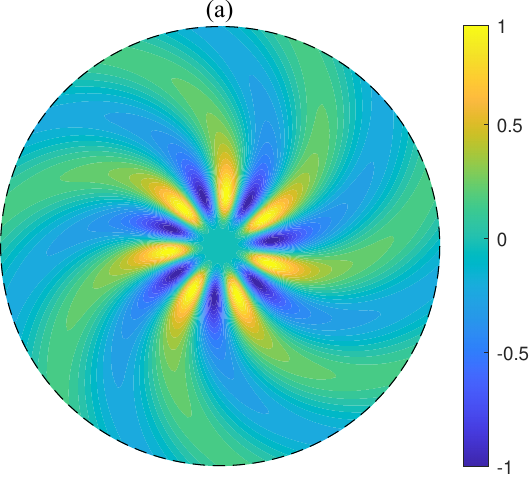}%
    \hspace{1cm}
    \includegraphics[width=5cm]{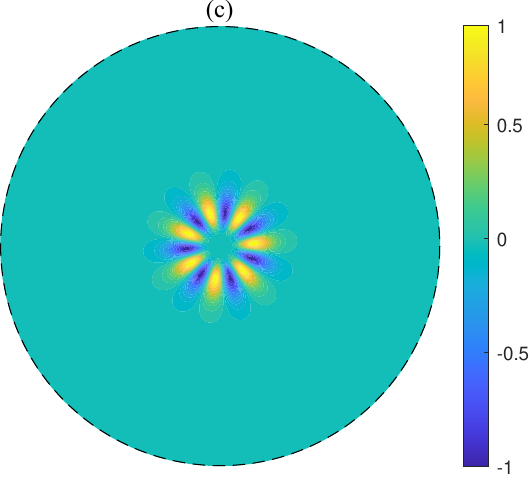}%
    \par\vspace{.5cm}
    \includegraphics[width=5cm]{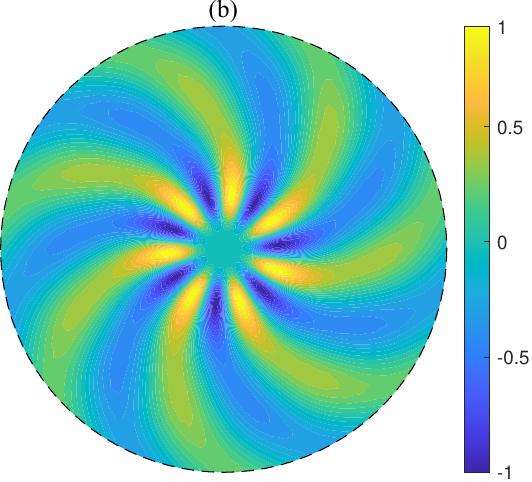}%
    \hspace{1cm}
    \includegraphics[width=5cm]{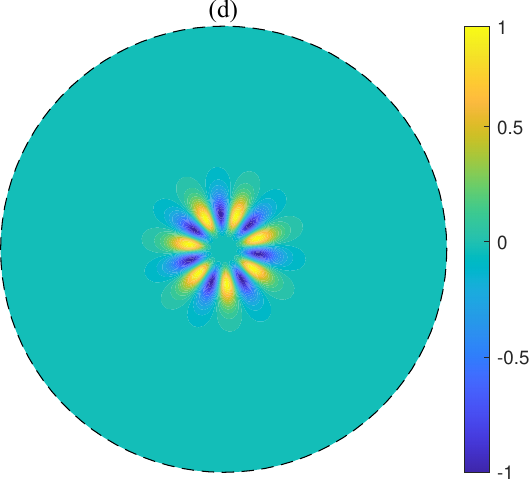}%
    \par\vspace{.75cm}
    \includegraphics[width=5cm, height=3cm]{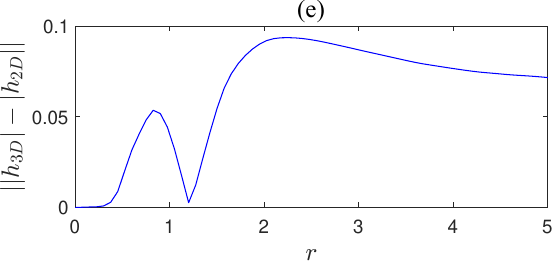}%
    \hspace{1cm}
    \includegraphics[width=5cm, height=3cm]{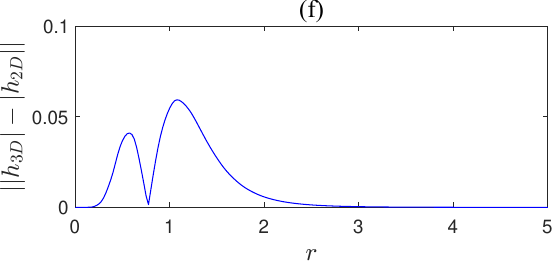}%
\caption{Comparison of perturbation height eigenmodes between the 3D computations and our reduced model. (a): radiating mode $(m = 7, F = 0.3, \hinfty = 1, \omega = -1.377 - 0.014\I)$ from 3D numerics. (b): radiating mode $(m = 7, F = 0.3, \omega = -1.374 - 0.013\I)$ from model. (c): trapped mode $(m = 7, F = 0.5, \hinfty = 1, \omega = -0.536-10^{-9}\I)$ from 3D numerics. (d): trapped mode $(m = 7, F = 0.5, \omega = -0.540-10^{-13}\I)$ from model.
(e): norm of the error between radiating mode from 3D numerics ($|h_{3D}|$) and model ($|h_{2D}|$).
(f): norm of the error between trapped mode from 3D numerics ($|h_{3D}|$) and model ($|h_{2D}|$).}
\label{fig:commparison_3D_model_from_above}
\end{figure}
\begin{figure}
    \centering%
        \includegraphics[width=7cm]{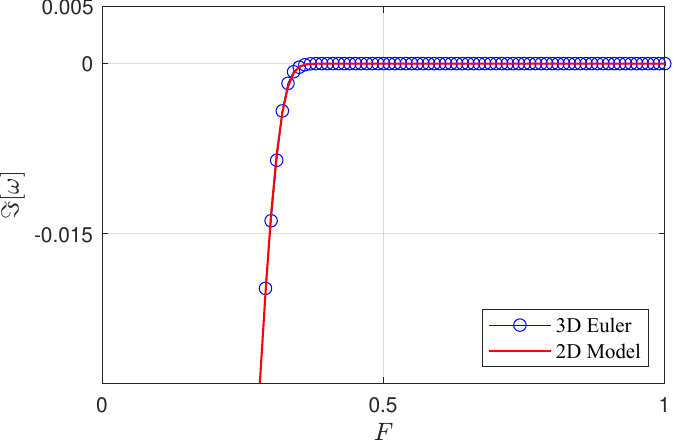}%
        \includegraphics[width=7cm]{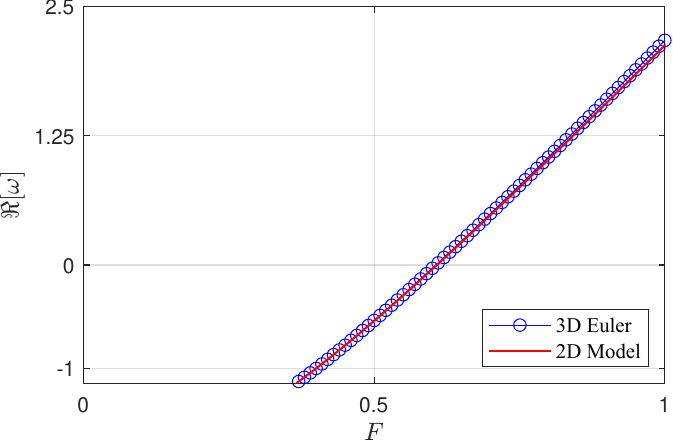}
    \caption{Comparison of the least damped modes spectrum as function of the Froude number for the Lamb-Oseen vortex flow between the 3D simulations and the model. Parameters have been taken as $m=7$ and $\hinfty = 5$.
    }
    \label{fig:eigenvalues_and_eigenfunction_comparison}
\end{figure}

%
In figure~\ref{fig:commparison_3D_model_from_above} we show a comparison between the eigenmodes from the Euler and model equations for a radiating (at $F=0.3$) and trapped (at $F=0.5)$ mode with azimuthal wavenumber $m=7$, and in the case of the Euler equations, height $\hinfty = 1$. 
Here we show the modes in the horizontal plane (``seen from above''). We plot the pressure, or equivalently the height fields. The modes are complex, corresponding to rotating waves, and the phases plotted in figure 
\ref{fig:commparison_3D_model_from_above} are arbitrary. The modes from the model are nearly identical with those from the Euler equations. This is expected given that we have verified the accuracy of the closure condition in these cases. 
A more quantitative comparison between the two modes is given in panels (e) and (f), which show the norm of the difference in the modulus of the free surface height $\big||h_\mathrm{3D}|-|h_\mathrm{2D}|\big|$ as functions of the radial coordinate for the radiating mode ($F = 0.3$) and the trapped mode ($F = 0.5$), respectively. The absolute error is at most $0.1$ in the radiating case and about $0.05$ for the trapped mode case, confirming the accuracy of the model predictions.
In addition, a direct comparison of the eigenvalues as function of Froude number computed using 3D numerics and model is given in figure \ref{fig:eigenvalues_and_eigenfunction_comparison}.
The agreement is excellent even at moderately large Froude numbers where the surface deformation is considerable.

\section{Conclusion}\label{sec:conclusions}

In this paper we have presented a novel, spatially two-dimensional set of equations to study the propagation of deep-water surface waves over a general steady base vortex flow. The model overcomes the widely used approximations of neglecting the free surface deformation and of considering potential base flows. The model is obtained by evaluating the linearized Euler equations on the base free surface and then introducing a closure equation to account for the vertical derivatives at the free surface. 
The model has been derived from a heuristic argument, with numerical results used to validate the closure. Specifically, we have computed eigenmodes of the fully 3D Euler equations for the Lamb-Oseen vortex with various fluid depths and have directly evaluated the closure condition for these modes. In the deep-water regime, the closure condition is found to hold. We have verified this for both radiating and trapped modes over the Lamb-Oseen vortex and we have extended this test for two other vortex profiles. Furthermore, we have directly compared leading eigenvalues and eigenmodes from the full 3D computations and the model, and the agreement is excellent over a wide range of Froude numbers. While we have presented only results for specific parameters, e.g., azimuthal mode number $m = 7$, we have found these results to be representative of a variety of other conditions. The overall comparison is excellent and confirm the accuracy of our reduced model.

More than just reducing the number of spatial dimensions, the model presented allows for an exact non-reflecting boundary condition in the frequency domain (as derived in appendix~\ref{appendix:model_NRBC}). This is important for simulations of infinitely extended waves and radiating eigenmodes in finite computational domains. For the full 3D Euler equations, such exact non-reflecting boundary conditions are not available, and numerically preventing reflections, for example by using an absorbing layer as in~\citep{zuccoli+brambley+barkley-2022}, further increases the computational demand.

The closure condition used to recover our two-dimensional model is only valid for surface gravity waves, and we have shown that other waves such as inertial waves are not captured by it.  This is both a blessing and a curse: by using our two-dimensional model, we are able to calculate only the surface gravity waves without needing to waste computation on computing other modes; however, the other modes are unobtainable using our method. We note that we do not find spurious modes that are an artifact of the closure condition. These observations may be of use when considering how a rigorous derivation of our condition may be obtained, as at some point in the derivation an assumption specific to surface gravity waves must be made.

As future work we leave the possibility of formally justifying the closure condition, using the model to investigate the propagation of waves over axisymmetric vortices, and generalizing the model to non-axisymmetric base flows. While we have attempted to justify the model closure~\citep{zuccoli-2023}, we have been unable to find any argument or formal calculation justifying it and hence we leave this for future work; the obvious assumption of a simple exponential decay with depth away from the free surface fails due to significant radial variation of the exponential decay rate, and more sophisticated multiple-scales-like analyses would rely on a separation of length scales not apparent in figure~\ref{fig:closure_radiating}, for example.
We have validated the model against eigenmodes of the full Euler equations because these can be obtained at moderate computational cost in deep water owing to separability in the azimuthal direction. However, the model is a time-dependent, two-dimensional system that is not restricted to describing just eigenmode solutions. We envision it being used to investigate phenomena that would be extremely challenging to simulate in the full Euler equations,
for example, the scattering of surface waves by strong vortices in deep water.
Generalizing the closure to more general base flows, i.e.\ non-axisymmetric and involving three velocity components, appears to be possible, but introduces a lot of complexity that we have opted not to address here. 
A fully general model could be used to make predictions of the behaviour of deep-water surface waves on complex flows with complex surface deformations that are out of reach with fully three-dimensional simulations, such as a vortex dipole, as briefly discussed by \citet{vivanco2004experimental} from an experimental point of view.

\begin{acknowledgments}
E.~Zuccoli was funded through the Warwick Mathematics Institute Centre for Doctoral Training, and gratefully acknowledges the support of the University of Warwick and the UK Engineering and Physical Sciences Research Council (EPSRC grant EP/W523793/1).
E.J.~Brambley gratefully acknowledges the support of the UK Engineering and Physical Sciences Research Council (EPSRC grant EP/V002929/1).
D.~Barkley gratefully acknowledges supported from the Simons Foundation (Grant No. 662985).
\end{acknowledgments}

\appendix  

\section{Closure for the flat, 2D Cartesian case}
\label{appendix:2D_closure}

Insight into our closure condition can be gained by examining the corresponding closure for the simple case of surface waves on a 2D fluid at rest, and then considering the formal generalization to a swirling base flow.  
Let the depth of the 2D fluid be infinite and take the flat free surface to be at $z=0$. Then the governing equations for linear perturbation are the horizontal and vertical momentum balances and incompressibility, together with the kinematic and dynamic boundary conditions (BCs): 
\begin{subequations}\begin{align} &
    \frac{\partial u_x}{\partial t}
 + \frac{1}{\rho}\frac{\partial p}{\partial x} = 0,
 \qquad
    \frac{\partial u_z}{\partial t}
 + \frac{1}{\rho}\frac{\partial p}{\partial z} = 0,
\qquad
\frac{\partial u_x}{\partial x} + \frac{\partial u_z}{\partial z} = 0, 
\\ &
\frac{\partial h}{\partial t} - u_{z} = 0, 
\qquad \qquad
p - \rho gh = 0, \quad {\rm on} \quad z = 0. \label{eq:2DpBC}
\end{align}\end{subequations}%
For surface waves, additionally one supposes that all perturbation fields vanish as $z \to -\infty$.

We substitute the dynamic BC into the kinematic BC to eliminate $h$ in favor of $p$. We then invoke the ansatz of this paper: we assume that the derivative with respect to $z$ of this expression vanishes at the free surface, giving
\begin{equation}
    \frac{\partial u_z}{\partial z} = \frac{1}{\rho g} \frac{\partial}{\partial t} 
    \Big( \frac{\partial p}{\partial z} \Big), \quad {\rm on} \quad z = 0.
    \label{eq:2Dclosure1}
\end{equation} 
This expression clearly holds because surface-wave solutions in this case are easily found by separation and vary in $z$ with dependence $e^{|k|z}$, for real $k$. We put this aside and continue the formal development. Using vertical momentum balance, Eq.~\eqref{eq:2Dclosure1} becomes
\begin{equation}
    \frac{\partial u_z}{\partial z} = -\frac{1}{g} \frac{\partial^2 u_z}{\partial t^2} , \quad {\rm on} \quad z = 0.
    \label{eq:2Dclosure2}
\end{equation} 
Using the kinematic BC to express $u_z$ at the free surface in terms of $h$, this takes the final form
\begin{equation}
    \frac{\partial u_z}{\partial z} = -\frac{1}{g} \frac{\partial^3 h}{\partial t^3}, \quad {\rm on} \quad z = 0.
    \label{eq:2Dclosure3}
\end{equation} 
Equations \eqref{eq:2Dclosure1}-\eqref{eq:2Dclosure3} are the 2D Cartesian equivalent of the closure Eqs.~\eqref{eq:closure_BC_all}. The final form, Eq.~\eqref{eq:2Dclosure3}, is what we use to close the remaining equations. 

The remaining equations are the horizontal momentum and the incompressibility constraint. We evaluate these on the free surface, using the dynamic BC to eliminate $p$ in terms of $h$ in the horizontal momentum equation and using closure \eqref{eq:2Dclosure3} to eliminate $\partial u_z/ \partial z$ in the incompressibility constraint. Letting $\displaystyle{u = u_x |_{z=0}}$, we have the closed equations on the free surface
\begin{align}
    \frac{\partial u}{\partial t} + g \frac{\partial h}{\partial x} &= 0, &
\frac{\partial u}{\partial x} -\frac{1}{g} \frac{\partial^3 h}{\partial t^3} &= 0. \label{eq:2Dfinal}
\end{align} 
These are the 2D Cartesian equivalent of Eqs.~\eqref{final_reduced_system}. 

As already noted, the validity of all the above for linear surface waves is readily obvious from the known form of such waves. 
Equation~\eqref{eq:2Dclosure2} clearly holds for $u_z \sim e^{i \omega t - ikx} e^{|k|z}$ with the dispersion relation $\omega^2 = g |k|$. Equations \eqref{eq:2Dfinal} can be combined to give
\begin{equation}
\frac{\partial^4 h}{\partial t^4} = - g^2 \frac{\partial^2 h}{\partial x^2},
\end{equation} 
which is satisfied for surface waves $h \sim e^{i \omega t - ikx}$ with the dispersion relation $\omega^2 = g |k|$. 

The formulation in Sec.~\ref{sec:maths_model} generalizes these statements to the case of a swirling flow with a deformed free surface. Specifically, Eq.~\eqref{eq:2Dclosure2} generalizes to
\begin{equation}
     \frac{\partial v_s}{\partial z} = -\frac{1}{g} D^2_t u_z
    \quad {\rm on} \quad z = \hzero(r),
    \label{eq:formal_BC}
\end{equation} 
upon formal replacement  $u_z \to v_s = u_{z} - \hzeroprime u_{r}$ on the LHS and $\partial/\partial t \to D_t$ on the RHS. The vertical velocity on the LHS of Eq.~\eqref{eq:2Dclosure2} comes from the vertical motion of the free surface via the kinematic boundary condition.
The generalization of this quantity to the deformed free surface is $v_s$, since $v_s(r,\hzero(r))$ is the vertical velocity of the free surface. The vertical velocity on the RHS of Eq.~\eqref{eq:2Dclosure2} comes from the vertical momentum and remains vertical momentum for the non-flat case. 
Equation~\eqref{eq:formal_BC} is closure condition \eqref{eq:closure_BC_w}.

This suggests also an alternative view of the model derivation. We may simply postulate that \eqref{eq:formal_BC} holds, as the natural generalization of Eq.~\eqref{eq:2Dclosure2} to the case of swirling base flows.   
This assumption alone establishes the model closure via the equalities in Eqs.~\eqref{eq:closure_BC_all}.

\section{Exact non-reflecting boundary condition for a single frequency}
\label{appendix:model_NRBC}

When the base flow is a radially decaying vortex and the perturbation has a single frequency $\omega$, as all testing cases considered here are, the reduced model equations (\ref{final_system_single_vortex_dimensionless}) allow for the imposition of an exact non-reflecting boundary condition at spatial infinity, similarly to the shallow-water case.
In order to show this, let us consider the equations in the far-field limit. Assuming a time dependence $\exp\{-\I\omega t\}$, as the base flow vanishes, the set of equations~\eqref{final_system_single_vortex_dimensionless} becomes
\begin{subequations}
\label{final_system_single_vortex_dimensionless_far_field}
    \begin{align} &
    -\I\omega\vect{u}+ \bnablaperp h = 0, \\ &
    -\I\omega^3h + \vect{\nablaperp\cdot u} = 0,
    \end{align}
\end{subequations}
where $\vect{u} = (u, v)$ is the horizontal velocity field on the free surface. The equations above can be combined into a single equation for the perturbation height, reading
\begin{equation}
        \nablaperp^{2} h + \omega^4 h = 0.
    \label{equ:single_eq_h_far_field}
\end{equation}
This is Helmholtz equation, and so the solution may be written using Hankel functions,
\begin{equation}
h(r,\theta,t) = \sum_{m=-\infty}^\infty \Big(a_mH_m^{(1)}(\omega^2r) + b_mH_m^{(2)}(\omega^2r)\Big)\exp\{-\I\omega t + \I m\theta\}
\label{equ:hankel}
\end{equation}
Recalling the asymptotic behaviour of Hankel functions in the far field~\citep{abra-1965}, equation~\eqref{equ:hankel} in the limit $r\rightarrow\infty$ behaves as
\begin{equation}
h(r,\theta,t) \sim \sum_{m=-\infty}^\infty \left(
    \frac{A_m}{\sqrt{r}}\exp\{-\I\omega t + \I\omega^2r + \I m\theta\}
  + \frac{B_m}{\sqrt{r}}\exp\{-\I\omega t - \I\omega^2r + \I m\theta\}
  \right),
    \label{equ:asympt_sol_h}
\end{equation}
where the asymptotic coefficients are given in terms of the Hankel function coefficients by
\begin{align}
A_m &= a_m\sqrt{\frac{2}{\pi\omega^2}}\exp\{-\I (2m+1)\pi/4\}, &
B_m &= b_m\sqrt{\frac{2}{\pi\omega^2}}\exp\{ \I (2m+1)\pi/4\}.
\end{align}
Not only does equation~\eqref{equ:asympt_sol_h} return the exact dispersion relation for deep-water gravity waves when the base flow is negligible ($\omega^2 = k$, recalling that we have nondimensionalized such that $g=1$), but it also represents the far field solution as outgoing waves (with coefficients $A_m$) and incoming waves (with coefficients $B_m$), provided $\Real(\omega)>0$. A non-reflecting boundary condition would then imply that the incoming $B_m$ coefficients are zero independently of the outgoing $A_m$ coefficients.  However, due to the dispersive nature of such waves, the distinction between incoming and outgoing waves is more subtle here and is reversed when $\Real(\omega)<0$, such that in that case $A_m$ are the incoming coefficients and $B_m$ are the outgoing coefficients.  This leads to the following two exact non-reflecting boundary conditions in the far field:
\begin{enumerate}
    \item If $\Real(\omega) > 0$, the exact NRBC reads:
    \begin{equation}
        \begin{aligned} &
            -\I\omega^2\Big(\sqrt{r}h \Big) + \frac{\partial}{\partial r}\Big( \sqrt{r}h \Big) = 0, \quad r \rightarrow\infty.
        \end{aligned}
    \label{equ:exact_NRBC_plus_sign_omega}
    \end{equation}
    \item Viceversa, if $\Real(\omega) < 0$, the exact NRBC reads
    \begin{equation}
        \begin{aligned} &
            +\I\omega^2\Big( \sqrt{r}h \Big) + \frac{\partial}{\partial r}\Big( \sqrt{r}h \Big) = 0, \quad r \rightarrow\infty.
        \end{aligned}
    \label{equ:exact_NRBC_minus_sign_omega}
    \end{equation}
\end{enumerate}
If, as above, the frequency $\omega$ is the eigenvalue of the problem being solved, one numerical strategy would be to implement two separate codes; the first code would implement boundary condition \eqref{equ:exact_NRBC_plus_sign_omega} and retain only modes whose frequencies have a positive real part, while the second code would implement boundary condition \eqref{equ:exact_NRBC_minus_sign_omega} and retain only modes whose frequencies have a negative real part. The full solution is then the union of these two sets of modes.

\bibliography{bibliography}

\end{document}